\documentclass[12pt]{article}
\usepackage{epsf}
\usepackage{graphicx}
\usepackage{eepic}
\usepackage{latexsym}
\newcommand{\be}{\begin{equation}}
\newcommand{\ee}{\end{equation}}
\newcommand{\ba}{\begin{array}}
\newcommand{\ea}{\end{array}}
\newcommand{\bqa}{\begin{eqnarray}}
\newcommand{\eqa}{\end{eqnarray}}
\begin{document}

\begin{center}
{\Large\bf Is the $f_0(600)$ meson a dynamically generated
resonance? -- a lesson learned from  the O(N)  model and beyond }
\\[10mm]
{\sc L.~Y.~Xiao, Zhi-Hui Guo and H.~Q.~Zheng}
\\[2mm]
{\it  Department of Physics, Peking University, Beijing 100871,
P.~R.~China}
\\[5mm]
\today
\end{center}

\begin{abstract}
 O(N) linear $\sigma$ model is solvable in the large $N$ limit and
 hence provides a useful theoretical laboratory to test various
 unitarization approximations. We find that the large $N_c$ limit and the
 $m_\sigma\to \infty$ limit do not commute. In order to get the correct large $N_c$ spectrum
 one has to firstly take the large $N_c$ limit. We argue that the $f_0(600)$ meson
may not be  described as generated dynamically. On the contrary, it
 is most appropriately described at the same level as the pions, i.e, both
 appear explicitly in the effective lagrangian.
Actually it is very likely the $\sigma$ meson responsible for the
spontaneous chiral symmetry breaking in a lagrangian with linearly
realized chiral symmetry.
\end{abstract}

There have been remarkable progresses in recent  years in the field
of low energy strong interaction physics. In the I=0, J=0 channel of
 $\pi\pi$ scatterings,  it is demonstrated that the $f_0(600)$ (or $\sigma$)
 meson is crucial to
adjust chiral perturbation theory to experiments~\cite{XZ00}, and
in fact, the $f_0(600)$ meson provides the dominant contribution
to the phase shift at low energies. The Roy type equation
analyses~\cite{CCL}  clearly indicate the existence of a light and
broad $S$ matrix pole in the $IJ=00$ channel of $\pi\pi$ elastic
scattering amplitude, and also in the I=1/2, J=0 channel $\pi K$
scattering amplitude ($K^*(800)$ or $\kappa$)~\cite{DM06}. The
dispersive analyses of Refs.~\cite{PKU1,PKU2,PKU3}, firmly
demonstrate the existence of the $\sigma$ and the $\kappa$
resonances as well. All the results on the pole positions based on
dispersion techniques agree with each other rather well
convincingly, within error bars -- hence, in the authors' opinion,
settling down the long debate on whether there exist such light
and broad resonances. The $\sigma$ meson can also be understood in
other approaches. For example, using Dyson--Schwinger equations,
in Ref.~\cite{Roberts} the mass and even the large width of the
$\sigma$ meson can be determined rather well.

Having firmly established the existence of sigma and kappa, the next
question would be what these resonances are. There exists a long
list of references with quite different opinions. For example, there
are attempts to explain these light and broad resonances in linear
sigma models at hadron level~\cite{lsigmam} or at quark
level~\cite{scadron}. Interestingly, it was observed that  poles
corresponding to these lightest scalar resonances can be found in
the unitarized amplitudes calculated using chiral perturbation
theory, i.e., not having to include the scalars explicitly in the
effective lagrangian. Hence the lightest scalars are considered as
`dynamically generated'~\cite{pelaez}. Unlike a typical resonance,
for example, the $\rho(770)$ meson, it is found that the $\sigma$
and $\kappa$  resonance poles move to infinity on the complex $s$
plane in the large $N_c$ limit~\cite{pelaez2}. This different $N_c$
behavior is hence believed to be an evidence to support that the
$\sigma$ and $\kappa$ are generated dynamically. This aim of this
paper is to examine the latter approach closely and try to clarify
the meaning of `dynamically generated' resonance.

The idea that a particle can be generated dynamically is not new.
Basdevant and  Lee have succeeded in generating a dynamical $\rho$
resonance from the renormalizable linear sigma
model~\cite{basdevant}. On the other side, Basdevant and Zinn-Justin
studied the massive Yang-Mills lagrangian in which the $\rho$ meson
acts as the massive gauge boson. Through a proper unitarization
procedure, they generate in  $d$ wave the $f_2(1270)$ resonance and
a broad $s$ wave $\sigma$ resonance~\cite{basdevant2}. These
activities are closely connected to the old concept of bootstrap
duality: in one lagrangian model, a particle may be called
elementary (i.e., exhibits itself explicitly in the lagrangian),
others are called dynamical  (i.e., only show up as $S$ matrix poles
but not included in the lagrangian); whereas in other lagrangian
models, the particle's role as `elementary' or `dynamical' can be
just reversed.

 The idea of bootstrap duality is very appealing.
However it is difficult to justify it in practice, since it involves
nonperturbative calculations of field theory. The only way at
present is to use unitarization approximations to investigate it.
One of the most frequently used method is the Pad\'e approximation.
Nevertheless it is difficult to justify to what extent one can trust
the predictions of Pad\'e approximation, especially when it is
applied to chiral perturbation theory ($\chi$PT)
amplitudes~\cite{qin02}. Concerning the difficulty of the problem,
in this letter we will firstly focus on  the $O(N)$ linear sigma
model. This model is solvable in the large $N$
limit~\cite{coleman74} and hence provides a good theoretical
laboratory to test the validity of various unitarization
approximations.
From these studies, one can gain better
understandings to the unitarization approximation, when it may be
trusted and to what extent they can be trusted.

The lagrangian of linear O(N) model is,
 \be\label{lag1}
\mathcal{L}=\frac{1}{2}\partial_\mu\Phi^T\partial^\mu\Phi-\frac{1}{2}m^2\Phi^T\Phi-\frac{\lambda}{8N}(\Phi^T\Phi)^2
\ee
 where $\Phi=(\Phi_1,\Phi_2,\cdots,\Phi_N)^T$.
  We work in the symmetry breaking phase, so $m^2<0$ and the $\Phi$
  field develops a VEV,
$v^2\equiv<\Phi^T\Phi>=-2Nm^2/\lambda$ which is proportional to N
(i.e., number of flavors).
  To be more realistic, we discuss the $\pi\pi$
scattering in O(N) model with massive pions, by adding an explicit
symmetry breaking lagrangian to the above $\cal L$,
 \be\label{Lsb}
  {\cal L}_{S.B.}=c\Phi_N
   \ee
and $c=vm_\pi^2$.
 The
pseudo-goldstone boson scattering amplitude is,
 \be
T(\pi_a\pi_b\rightarrow\pi_c\pi_d)=iD(s)\delta_{ab}\delta_{cd}+iD(t)\delta_{ac}\delta_{bd}+
iD(u)\delta_{ad}\delta_{bc}\ , \ee
  where
 \bqa\label{D-1}
  D^{-1}(s)=-i\left(
              \begin{array}{cc}
               s-m^2_\pi & -v  \\
                -v & \frac{N}{\lambda}+ N B_0(s;m_\pi^2)
              \end{array}
            \right)\ ,
\eqa
 and also
 \be\label{B0}
B_0(p^2,m_\pi^2)=-\frac{i}{2}\int\frac{d^Dk}{(2\pi)^D}\frac{1}{(k^2-m_\pi^2)((k+p)^2-m_\pi^2)}\
.
 \ee
  The integral in above is made finite using dimensional
regularization,
 \bqa
&&\int\frac{d^Dk}{(2\pi)^D}\frac{1}{k^2-m^2_\pi}\frac{1}{(p+k)^2-m^2_\pi}=i\Delta-\frac{i}{16\pi^2}\widetilde{B}_0(s,m^2_\pi)\
,
 \eqa
 with ($\epsilon=2-D/2$)
 \be
  \Delta=\frac{1}{16\pi^2}[\Gamma(\epsilon)+\ln{4\pi}+\ln{\frac{\nu^2}{m^2_\pi}}]\ ,
   \ee
 \be
\widetilde{B}_0(s,m^2_\pi)=-2-i\sqrt{\frac{4m^2_\pi-s}{s}}(\ln{\frac{1-i\sqrt{\frac{4m^2_\pi-s}{s}}}
{1+i\sqrt{\frac{4m^2_\pi-s}{s}}}}+i\pi)\ .
 \ee
  In order to compare with
the calculations made later in this paper, notice that here we adopt
a different strategy to the $O(N)$ model: we treat it as a cutoff
effective theory, which means all the parameters in lagrangian are
taken as finite. In cutoff effective theory one introduces a cutoff
$\Lambda$ in the above divergent integral. This can also be done
using dimensional regularization scheme, by introducing the cutoff
$\Lambda$ in the following manner,
 \be
\Gamma(\epsilon)+\ln{4\pi}+\ln{\frac{\nu^2}{m_\pi^2}}\Rightarrow
\ln{\frac{\Lambda^2}{m_\pi^2}}\ .
 \ee
One can demonstrate that the above procedure is physically
equivalent to using a sharp momentum cutoff method, when the energy
scale is well below $\Lambda$. Notice that, as clearly shown from
Eqs.~(\ref{D-1}) and (\ref{B0}) that the notorious square divergence
does not occur in the scattering amplitude. The reason is that the
square divergence only occurs in the effective potential, and once
it is absorbed by the bare $\sigma$ mass through mass
renormalization it disappears entirely everywhere else. This
observation justifies the negligence of square divergence in the
latter calculation in effective theory. Another issue with respect
to the $O(N)$ model is that  the $O(N)$ model may contain a tachyon
in some regularization schemes. The tachyon also appears in the
present cutoff version. But it is numerically verified that the
tachyon pole locates at a scale much larger than the cutoff itself
and is hence of no concern. In the cutoff version the parameter
$\lambda$ appeared in Eq.~(\ref{D-1}) is the same as the bare
parameter $\lambda$ appeared in the lagrangian Eq.~(\ref{lag1}) and
obeys the relation
 \be
  \lambda=\frac{m_\sigma^2-m_\pi^2}{v^2/N}\ .
 \ee

The denominator of the propagator $D$ is given by the following
expression,
 \be G\equiv(s-m^2_\pi)\{{1\over\lambda}-{1\over
32\pi^2}[{\widetilde B}_0(s,m_\pi^2)+\ln{m^2_\pi\over \Lambda^2}]\}-
\frac{v^2}{N}\ .
\ee%
Poles appeared in $\pi\pi$ scattering are determined by solving the
equation $G^{\rm II}=0$, where $G^{\rm II}$ is the analytic
continuation of $G$ on the second sheet. From above we realize that
the true sigma pole position is a function of the bare sigma mass
parameter, for fixed cutoff $\Lambda$.  The sigma pole trajectory is
plotted in Fig.~1. Throughout this paper we take
$v/\sqrt{N}=f_\pi=93$MeV and $\Lambda=1.5$GeV. The bare mass of
sigma in Fig.~1 ranges from 400MeV to 1200MeV, and each value
differs by 100MeV.

In the following we will discuss variations of the above O(N)
lagrangian and calculate the scattering amplitude using different
unitarization approximations. By comparing with the above rigorous
results, we can gain some knowledge on the quality of various
approximation methods. It is worth noticing that, for the toy linear
O(N) model, it is demonstrated that the [n,n] Pad\'e amplitudes
gives the exact sigma pole location, as comparing with the exact
solution~\cite{willenbroch}. The K matrix approach does not
reproduce the exact sigma pole location, but it is still a good
approximation~\cite{willenbroch}. In the following we will show that
the nice property of these approximation methods will not be
maintained if the pion fields are expressed in the non-linear
representation.

The next thing we will do is to make polar decomposition of the O(N)
field, $i.e.$,
 \be
\Phi\equiv\varphi\vec{\phi},\,\,\,\phi=(\phi_1,\phi_2,......\phi_N)
 \ee
  with
constraint
 \be\label{cons}
  \phi_1^2+\phi_2^2+\phi_3^2+.......+\phi_N^2=1\ .
  \ee
In the symmetry breaking phase we define the shifted field,
 \be
  \varphi=\sigma + v
 \ee
and normalize the field $\vec{\phi}$ as
 \be
  \vec{\pi}=v\vec{\phi}\ .
  \ee
   With above
preparations we can obtain the polar decomposition lagrangian,
 \be\label{polar}
\mathcal{L}_{polar}=\mathcal{L}^{\sigma}+\frac{1}{2}(1+\frac{\sigma}{v})^2(\partial_\mu\vec{\pi}\cdot\partial^\mu\vec{\pi}
+\partial_\mu\sqrt{v^2-\vec{\pi}\cdot\vec{\pi}}\partial^\mu\sqrt{v^2-\vec{\pi}\cdot\vec{\pi}})
\ee
 with
  \be
\mathcal{L}^{\sigma}=\frac{1}{2}\partial_\mu\sigma\partial^\mu\sigma-\frac{1}{2}m^2(\sigma+v)^2-\frac{\lambda}{8N}(\sigma+v)^4\
.
 \ee
 In the above formula we have used the constraint condition,
Eq.~(\ref{cons}), so $\vec{\pi}=(\pi_1,\pi_2,.....\pi_{N-1})$. Now
the symmetry breaking lagrangian, Eq.~(\ref{Lsb}) reads,
 \be
\mathcal{L}_{SB}=c(\sigma+v)\sqrt{1-\frac{\vec{\pi}\cdot\vec{\pi}}{v^2}}\
.
 \ee
Then we expand the square root to get the interacting vertices which
are needed for evaluating $O(p^4)$ $\pi\pi\to\pi\pi$ processes.
 \bqa\label{polar'}
 &&  \mathcal{L}_p=
\mathcal{L}^{\sigma}+\frac{1}{2}(1+\frac{\sigma}{v})^2(\partial_\mu\vec{\pi}\cdot\partial^\mu\vec{\pi}
+\frac{1}{v^2-\vec{\pi}\cdot\vec{\pi}}(\vec{\pi}\cdot\partial_\mu\vec{\pi})^2)+c(\sigma+v)\sqrt{1-\frac{\vec{\pi}\cdot\vec{\pi}}{v^2}}
 \nonumber \\&&
=
\mathcal{L}^{\sigma}+\frac{1}{2}\partial_\mu\vec{\pi}\cdot\partial^\mu\vec{\pi}+\frac{\sigma}{v}\partial_\mu\vec{\pi}\cdot\partial^\mu\vec{\pi}+
\frac{\sigma^2}{2v^2}\partial_\mu\vec{\pi}\cdot\partial^\mu\vec{\pi}+\frac{1}{2v^2}(\vec{\pi}\cdot\partial_\mu\vec{\pi})^2
\ \nonumber \\&&
+\frac{1}{2v^4}(\vec{\pi}\cdot\partial_\mu\vec{\pi})^2\vec{\pi}\cdot\vec{\pi}-\frac{m_{\pi}^2}{2}\vec{\pi}\cdot\vec{\pi}
-\frac{m_{\pi}^2}{2v}\sigma\vec{\pi}\cdot\vec{\pi}-\frac{m_{\pi}^2}{8v^2}(\vec{\pi}\cdot\vec{\pi})^2
-\frac{m_{\pi}^2}{16v^4}(\vec{\pi}\cdot\vec{\pi})^3 \nonumber\\
&&+\cdots\ .
 \eqa
 Notice that in the above lagrangian the $\sigma^2\pi^2$ term does
 not contribute to $\pi\pi$ scattering amplitude at 1-loop level
 since its contribution is 1/N suppressed.
  There are other variants of the above effective
lagrangian. For example we may simply neglect the $\sigma$ field in
Eq.~(\ref{polar}) and get the O(N) non-linear sigma model
 \be
\mathcal{L}_{O(N)\,
non-linear}=\frac{1}{2}(\partial_\mu\vec{\pi}\cdot\partial^\mu\vec{\pi}
+\partial_\mu\sqrt{v^2-\vec{\pi}\cdot\vec{\pi}}\partial^\mu\sqrt{v^2-\vec{\pi}\cdot\vec{\pi}})\
.
\ee
 Expanding the pion fields from the  above lagrangian one gets
(also obtainable from Eq.~(\ref{polar'}) by simply neglecting the
sigma field):
 \bqa\label{NL'}
 && \mathcal{L}_{NL}
=
\frac{1}{2}\partial_\mu\vec{\pi}\cdot\partial^\mu\vec{\pi}-\frac{m_{\pi}^2}{2}\vec{\pi}\cdot\vec{\pi}
 +\frac{1}{2v^2}(\vec{\pi}\cdot\partial_\mu\vec{\pi})^2
-\frac{m_{\pi}^2}{8v^2}(\vec{\pi}\cdot\vec{\pi})^2 \nonumber \\&&
+\frac{1}{2v^4}(\vec{\pi}\cdot\partial_\mu\vec{\pi})^2\vec{\pi}\cdot\vec{\pi}
-\frac{m_{\pi}^2}{16v^4}(\vec{\pi}\cdot\vec{\pi})^3  +\cdots\ .
 \eqa
Also one can integrate out the
 $\sigma$ field at tree level from Eq.~(\ref{polar}) to get a
 modified O(N) non-linear sigma model:
 \bqa\label{mod'}
\mathcal{L}_{\overline{NL}}=\mathcal{L}_{NL}+\frac{1}{2m_\sigma^2v^2}[(\partial_\mu\vec{\pi}\cdot\partial^\mu\vec{\pi})^2
-m_\pi^2\vec{\pi}\cdot\vec{\pi}\partial_\mu\vec{\pi}\cdot\partial^\mu\vec{\pi}+\frac{m_\pi^4}{4}(\vec{\pi}\cdot\vec{\pi})^2]
. \eqa This lagrangian resembles the $O(p^4)$ chiral perturbation
theory lagrangian.

Starting from  Eq.~(\ref{polar'}) we can calculate the
$\pi_a\pi_b\rightarrow\pi_c\pi_d$ scattering amplitude up to
$O(p^4)$ in large N limit:
\begin{enumerate}
 \item The contribution from tree level 4$\pi$ contact term, or current algebra term, denoted as
 $T_2$ ($\sim$$O(p^2))$,
 \bqa T_2
=
\delta_{ab}\delta_{cd}A_2(s)+\delta_{ac}\delta_{bd}A_2(t)+\delta_{ad}\delta_{bc}A_2(u)
\eqa
 where
\be A_2(s)=\frac{1}{v^2}(s-m_{\pi}^2) \ .\ee
\item The contribution from tree level $O(p^4)$ term contained in Eq.~(\ref{mod'}), denoted as
 $T_4$, is
 \bqa T_4
=
\delta_{ab}\delta_{cd}A_4(s)+\delta_{ac}\delta_{bd}A_4(t)+\delta_{ad}\delta_{bc}A_4(u)
\eqa
 where
\be A_4(s)=\frac{(s-m_{\pi}^2)^2}{m_\sigma^2v^2} \ .\ee
\item
The contribution from tree level $\sigma$ exchange in
Eq.~(\ref{polar'}) ($\sim O(p^4)$),
 \bqa
T_\sigma=\delta_{ab}\delta_{cd}A_\sigma(s)+\delta_{ac}\delta_{bd}A_\sigma(t)+\delta_{ad}\delta_{bc}A_\sigma(u)
\eqa
 where
 \be
  A_\sigma(s)=\frac{-
1}{v^2}\frac{(s-m_{\pi}^2)^2}{s-m_{\sigma}^2}\ ,
 \ee
 which is  $O(p^4)$ at low energies.
\item
The contribution from 1-loop diagrams (in large N limit and
including only $O(p^4)$ terms):
 \bqa
T_{loop}=\delta_{ab}\delta_{cd}A_{loop}(s)+\delta_{ac}\delta_{bd}A_{loop}(t)+\delta_{ad}\delta_{bc}A_{loop}(u)
\eqa
 where
  \bqa
   &&
A_{loop}=\frac{N}{v^4}[(s-m_{\pi}^2)^2B_0(s;m_\pi^2)-i\frac{3m_{\pi}^2-2s}{2}\int\frac{d^Dk}{(2\pi)^D}\frac{1}{k^2-m_{\pi}^2}] \ .
\nonumber\\
  \eqa
\end{enumerate}
The quadratic divergent integral from bubble diagram in above is
treated in the standard way in dimensional regularization
scheme,\footnote{If we were to use a square divergence, i.e., a
$\Lambda^2$ to replace the following integral, the resulting pole
location will be disastrous as comparing with the original $O(N)$
model prediction. Also a naive cutoff regularization scheme spoils
chiral symmetry and hence is not correct. An effort to remedy the
deficiency of the naive cutoff regularization is discussed in
Ref.~\cite{Wu}. }
 \be
\int\frac{d^Dk}{(2\pi)^D}\frac{1}{k^2-m^2}=im^2\Delta-\frac{i}{16\pi^2}\widetilde{A}(m^2)\
, \ee
 where
  \be \widetilde{A}(m^2)=-m^2\ .  \ee
 As before we introduce the cutoff in the same way as in calculating the solvable linear $O(N)$
 model, i.e.,
 $$
\Gamma(\epsilon)+\ln{4\pi}+\ln{\frac{\nu^2}{m^2}}\Rightarrow
\ln{\frac{\Lambda^2}{m^2}}\ .
 $$
One can check that the above treatment to ignore the square cutoff
leads to  the same result in the chiral limit to 1-loop amplitude
comparing with the one obtained from the original $O(N)$ model.

Now consider the I=0 channel,
 \be
  T^{I=0}(s,t,u)=(N-1)A(s)+A(t)+A(u)\ .
  \ee
   In the large N limit, only the $s$ channel amplitude $A(s)$ term survives.
   The J=0 partial wave is simply
    \be T^{00}=\frac{1}{32\pi}NA(s)\ . \ee

With the above preparation we will in the following construct
various unitarized  amplitudes based on tree and one loop
calculations using different lagrangians. The unitary amplitudes
being constructed are: $K_1$ (using tree level amplitude in pure
$O(N)$ non-linear $\sigma$ model), $K_2$ ($K_1$ plus $\sigma$
exchange), $K_3$  (1-loop calculation in the pure $O(N)$ non-linear
$\sigma$ model), $K_4$  (1-loop with $\sigma$ exchange included
using polar decomposed lagrangian); as well as $P_1$ ([1,1] Pad\'e
for the pure $O(N)$ non-linear $\sigma$ model), $P_2$ ([1,1] Pad\'e
with the $\sigma$ being integrated out, i.e., of Eq.~(\ref{mod'})),
$P_3$ ([1,1] Pad\'e with explicit $\sigma$ exchange, i.e., of
Eq.~(\ref{mod'})). For simplifying the expressions we in the
following use $v$ (=93MeV) to replace $v/\sqrt{N}$. In the following
we will always work in the IJ=00 channel and hence for simplicity we
drop out the superscripts 00 in the partial wave amplitude.

\paragraph{$K_1$ amplitude: } This
approximation corresponds to the simplest $K$ matrix amplitude in
the pure O(N) non-linear sigma model. Also it is the simplest $K$
matrix amplitude one obtains from Eq.~(\ref{polar'}). One has
 \be
S=\frac{1+i\rho K}{1-i\rho K}\ .
 \ee
Here $K_1=T_{2} $ and hence
 \be S=\frac{1+i\rho T_2 }{1-i\rho T_2 }\ ,
  \ee
  where $\rho=\sqrt{1-4m_\pi^2/s}$ and
  \be T_2 =\frac{1}{32\pi
v^2}(s-m_{\pi}^2) \ .
\ee
 The dynamically generated second sheet poles correspond to the zeros of the $S$ matrix,
 we show the pole position in Fig.~1
labeled as $K_1$. Notice that there is no free parameter here.  We
see from Fig.~1 that $K_1$ is not a good approximation numerically
to the real solution (i.e., the solution of the linear sigma model).
There is another serious disadvantage in this approximation, which
can be clearly seen in the chiral limit. That is the pole location,
$\sqrt{s_p}$, is proportional to $v$ and hence is proportional to
the square root of the number of colors, $N_c$:
 \be\label{solK_1}
  \sqrt{s_p}\propto \sqrt{i32\pi f_\pi^2}\propto \sqrt{N_c}\ .
\ee
 So when $N_c\to \infty$ this pole disappears! On the other side
the sigma pole in the linear sigma model behaves as $m_\sigma\sim
O(N_c^{0})$ in $N_c$ power counting.

\paragraph{$K_2$ amplitude:} $K_2=T_2+T_\sigma$.  This approximation
corresponds to calculating the scattering amplitude at tree level in
the polar decomposition lagrangian Eq.~(\ref{polar'}), and it is
also equivalent to the lowest order $K$ matrix amplitude evaluated
in the linear sigma model.\footnote{In the linear sigma model one
expands the amplitude according to the power of the coupling
constant $\lambda$ and constructs a $K$ matrix amplitude using the
tree level perturbative amplitude.} We have,
 \be
 S=\frac{1+i\rho (T_2+T_\sigma)}{1-i\rho (T_2+T_\sigma)}\ ,
  \ee
   with
\be T_\sigma=\frac{-1}{32\pi
v^2}\frac{(s-m_{\pi}^2)^2}{s-m_\sigma^2}\ ,
 \ee
 and $$K_2=T_2+T_\sigma= \frac{s-m_\pi^2}{32\pi
 v^2}\frac{m_\pi^2-m_\sigma^2}{s-m_\sigma^2}\ .$$
  The second sheet
poles correspond to the solution of the following equation,
 \be\label{solK_2}
s-m_\sigma^2+i\sqrt\frac{s-4m_\pi^2}{s}\frac{1}{32\pi
v^2}(m_{\pi}^2-m_\sigma^2)(s-m_{\pi}^2)=0\ .
 \ee
  The pole trajectory is
plotted in Fig.~1 labeled as $K_2$. Apparently $K_2$ works much
better than $K_1$. Especially the pole position is now
$\sqrt{s_p}\sim O(N_c^0)$. But numerically it is still a rather poor
approximation as can be seen from Fig.~1.

By analyzing the $K_2$ approximant one can learn an important
lesson. It is easy to find  solution of Eq.~(\ref{solK_2}) in the
chiral limit:
 \be\label{solK_2'}
  s_{pole}=\frac{m_\sigma^2}{1-\frac{im_\sigma^2}{32\pi v^2}}\ .
  \ee
In the above equation if one formally sends $m_\sigma\to\infty$ one
recovers the solution Eq.~(\ref{solK_1}) in the chiral limit,
$i.e.$, $s_{pole}=i32\pi v^2\to \infty $ when $N_c\to \infty$.
Nevertheless if one firstly take the large $N_c$ limit in
Eq.~(\ref{solK_2'}) one gets simply $s_{pole}=m_\sigma^2$.
Apparently the latter one is the correct relation we are looking
for. Therefore we conclude that \textit{one must take the large
$N_c$ limit before integrating out the sigma degree of freedom in
order to get the correct large $N_c$ spectrum.}

\paragraph{$K_3$ amplitude:} $K_3=T_2+{\rm Re}[T_{loop}]$. This
approximation corresponds to a 1--loop calculation in the pure O(N)
non-linear sigma model. Notice that here we adopt the cutoff
effective lagrangian approach, therefore  no renormalization
procedure and hence no counter terms are needed.
  \be
S=\frac{1+i\rho (T_2+{\rm Re}[T_{loop}])}{1-i\rho (T_2+{\rm
Re}[T_{loop}])} \ee with
 \bqa
&&  T_{loop}=\frac{1}{64\pi v^4}
\frac{1}{16\pi^2}\{(s-m_{\pi}^2)^2[2+\ln{\frac{\Lambda^2}{m_\pi^2}}+\rho(s)(\ln\frac{1-\rho(s)}{1+\rho(s)}+i\pi)]
\nonumber \\
&&+(3m_\pi^2-2s)m_\pi^2(1+\ln{\frac{\Lambda^2}{m_\pi^2}}) \}\ .
 \eqa
Notice that in this approximation the sigma pole is dynamically
generated and its location is also fixed. The pole position is shown
in Fig.~1 labeled as $K_3$. Notice that here, like the $K_1$ case,
one also has $\sqrt{s_p}\sim O(N_c^{1/2})$.

\paragraph{$K_4$ amplitude:} $K_4=T_2+T_\sigma+Re(T_{loop})$. This
approximation corresponds to considering all contributions up to
$O(p^4)$ term in the polar decomposed O(N) model. We have
 \be
S=\frac{1+i\rho [T_2+T_\sigma+Re(T_{loop})]}{1-i\rho
[T_2+T_\sigma+Re(T_{loop})]}\ . \ee
The pole trajectory is plotted in Fig.~1, labeled as $K_4$.
Comparing with $K_2$, the quality of the result is not actually
improved. What is even worse here is that the $K_4$ amplitude
contains a spurious pole on the second sheet, not far from the sigma
pole position.

We conclude in general that the $K$ matrix unitarization approach
 to the amplitude obtained from a lagrangian with non-linearly
 realized chiral symmetry does not work very well. Still, there exist
 other distant spurious poles (can be on both sheets)
 in these $K$ matrix amplitudes, which are not shown in Fig.~1.
  On the contrary, the
 $K$ matrix approach works fairly well on the linear  O(N)
 model.

  In the following we turn to discuss the Pad\'e
 approximations which are very frequently used in
 phenomenological discussions. Notice that Pad\'e amplitudes also preserve one
 bad heritage of the original exact amplitude -- the tachyon pole. But as
 stated before the appearance of the tachyon pole  does not matter in practice in the phenomenological discussions.
\paragraph{$P_1$ amplitude:} This approximation corresponds to the
[1,1] Pad\'e approximation in the cutoff pure non-linear O(N) model,
 \be
 T=\frac{T_2^2}{T_2-T_{loop}}\ .
  \ee
 The pole position is shown in Fig.~2 labeled as $P_1$. Certainly it is
 better than $K_1$ but similar in quality with $K_3$. The same
 problem remains, however, that is $\sqrt{s_p}\propto   v\propto
 O(N_c^{1/2})$.

\paragraph{$P_2$ amplitude:} This approximant corresponds to a 1-loop calculation
using $\cal{L}_{\overline{NL}}$ of Eq.~(\ref{mod'}):
 \be T=\frac{T_2^2}{T_2-T_4-T_{loop}}
  \ee
   where
    \be
T_4=\frac{(s-m_\pi^2)^2}{32\pi v^2m_\sigma^2}\ .
 \ee
Second sheet poles are determined by solving the following equation,
 \bqa\label{solp2}
&&s-m_\pi^2-\frac{(s-m_\pi^2)^2}{m_\sigma^2}-\frac{1}{32\pi^2v^2}\{(s-m_{\pi}^2)^2[2+\ln{\frac{\Lambda^2}{m_\pi^2}}
+\rho(\ln{\frac{1-\rho}{1+\rho}}-i\pi)]
\nonumber \\
&&+(3m_\pi^2-2s)m_\pi^2(1+\ln{\frac{\Lambda^2}{m_\pi^2}}) \}=0
 \eqa
The pole trajectory is shown in Fig.~2 labeled as $P_2$. This
approximation is the best among all. It is not difficult to examine
using Eq.~(\ref{solp2}) that if one takes the large $N_c$ limit one
finds $\sqrt{s_p}\simeq m_\sigma$. On the other sides, if one sends
$m_\sigma\to \infty$ before taking the large $N_c$ limit one gets
 $\sqrt{s_p}\propto v \sim O(N_c^{1/2})$. Once again we find that
 the $m_\sigma\to \infty$ limit and the large $N_c$ limit do not
 commute.

\paragraph{$P_3$ amplitude:} This approximation corresponds to the
[1,1] Pad\'e approximation up to $O(p^4)$ term in the polar
decomposition lagrangian Eq.~(\ref{polar'}). We have,
 \be
T=\frac{T_2^2}{T_2-T_\sigma-T_{loop}}\ .
 \ee
 The pole trajectory is plotted in Fig.~2 labeled as $P_3$. Notice
that adding a sigma pole explicitly in the lagrangian does not
improve the quality of output. What is even worse is that it also
predicts a spurious pole not very far from the sigma pole. Our
analysis seems to suggest that a polar decomposed sigma model
lagrangian works even worse than the situation when the sigma field
is completely integrated out. It is however not  clear whether this
observation remain valid in the realistic case.

The pole appeared in the unitarized amplitude is dynamical in the
case of $O(N)$ non-linear $\sigma$ model, and also dynamical in the
case of Eq.~(\ref{mod'}). In the former case the pole moves to
infinity as $N_c\to\infty$ whereas in the latter case the pole falls
down to the real axis when $N_c\to\infty$. For the latter we know
that the dynamical pole is nothing but just represents the original
$\sigma$ pole being integrated out from Eq.~(\ref{polar'}) or
Eq.~(\ref{lag1}), as is revealed by the simple exercise on $P_2$
amplitude, hence `elementary' in its origin. The drastically
different $N_c$ dependence of the pole trajectories, in our
understanding, merely reflects the different order when taking the
large $N_c$ and $m_\sigma\to\infty$ limits.

In a more realistic calculation however, the inverse problem becomes
problematic since the Pad\'e approximants for chiral perturbation
theory bring, together with the sigma pole, also spurious poles and
some of which violate analyticity~\cite{qin02}. Nevertheless the toy
model we discussed here well resembles many aspects of the real
situation. For example we pointed out that in the simplest $K_1$
approximant, the dynamical pole being generated has the deferent
$N_c$ behavior comparing with the $\sigma$ meson in the linear O(N)
model. The situation is very similar in realistic case. In
Ref.~\cite{leutwyler2}, based upon a [0,1] Pad\'e or $K_1$
approximant the authors suggest that the current algebra term
dominates at low energy, since the pole position of the sigma meson
obtained as such is,
 \be\label{ca}\sqrt{s_{\sigma}} \simeq
\sqrt{16i\pi f_\pi^2}\simeq 463-463i\ ,\ee
 and is found to be not far
from the pole position obtained from a more careful numerical
analysis. According to the analysis made in this paper, even though
the current algebra term dominates low energy physics, the $N_c$
dependence of the pole trajectory in Eq.~(\ref{ca}) does not in any
sense prove that the $\sigma$ pole is dynamical. The current algebra
term in Eq.~(\ref{polar'})\footnote{Or more precisely, when apply
$K_1$ approximation to Eq.~(\ref{polar'}).} gives exactly the same
$N_c$ dependence as in Eq.~(\ref{ca})\footnote{There is a factor of
2 difference between the coefficient in Eq.~(\ref{ca}) and
Eq.~(\ref{solK_1}).} meanwhile it comes from the $linear$ sigma
model with explicit $\sigma$ degree of of freedom! If Eq.~(\ref{ca})
were correct for arbitrarily
 large  $N_c$, the pole will move to infinity when $N_c\to \infty$,
 hence contradicting the $N_c$ counting rule for a conventional meson state.
 If this were true the sigma meson had to be a dynamical pole.
 Going beyond the leading current algebra amplitude, contrary to what is claimed in
Ref.~\cite{pelaez3}, in the [1,1] Pad\'e approximant in reality, the
$\sigma$ pole actually finally falls on the real axis. That is the
pole position as given in Eq.~(\ref{ca}) receive
 important corrections when $N_c$ is large:
  \bqa\label{11Pade}
&& s_{\sigma} = \frac{16i\pi f_\pi^2}{1+16i\pi f_\pi^2\triangle}\ ,\nonumber\\
&& \triangle=\frac{2}{3f_\pi^2}(22L_1+14L_2+11L_3)\propto O(N_c^0)\
,
 \eqa
 valid in the large $N_c$ and chiral limit. It is worth noticing that
it was not clear what is the approximation made to obtain
Eq.~(\ref{11Pade}).  Using the PKU parametrization
form~\cite{PKU1, PKU2,PKU3} one
 explicitly demonstrates that the Eq.~(\ref{11Pade}) is obtained
 under two further assumptions: 1) assuming one pole (the $\sigma$
 pole) dominates in the $s$ channel at low energies; 2) neglecting all the crossed
 channel resonance exchanges~\cite{sun}. It is understood that
 the assumption 1 actually fixes the  $N_c$ behavior of
 the $\sigma$ pole, i.e., $m_\sigma\sim O(1)$, $\Gamma_\sigma\sim
 O(1/N_c)$~\cite{XZ05,sun}. The assumption 2 does no harm in distorting
 the large $N_c$ behavior.
However,  these assumptions are only approximately correct
 when $N_c=3$ and fail at large  $N_c$, as demonstrated in
 Ref.~\cite{guo06}. Actually it is  likely that at large $N_c$, crossed channel
 resonance  exchange dominates low energy physics. Hence
 Eq.~(\ref{11Pade}) can neither be used to discuss the large $N_c$
 behavior of the $\sigma$ pole trajectory.
In Ref.~\cite{pelaez3}, it is observed that, unlike the $\rho$
meson which almost falls straightforwardly down to the real axis
when $N_c$ increases, the $\sigma$ meson is very unwilling to fall
down. Based on this observation the authors of Ref.~\cite{pelaez3}
concluded that the $\sigma$ pole they found from their unitarized
amplitude a dynamically generated one. Even if Eq.~(\ref{11Pade})
is invalid to be useful when discussing the large $N_c$ behavior
of the $\sigma$, the
 discussions given in this paper make it clear that the bent
 structure of the $\sigma$ pole trajectory should not be used
 to argue that the $\sigma$ is dynamical.~\footnote{We are of course fully aware of
 the fact that the toy $O(N)$ model is too simple to simulate QCD dynamics.} We plot in
Fig.~\ref{fig3} the $N_c$ dependence of the $\sigma$ pole in
linear $O(N)$ model (as well as the $P_2$ approximant).
Fig.~\ref{fig3} clearly shows that, even the `elementary' $\sigma$
does not fall down to the real axis straightforwardly. Though not
a rigorous proof, in our opinion the similar  bent structure of
the $\sigma$ pole trajectory both in Fig.~\ref{fig3} and in the
unitarized chiral perturbation theory amplitude suggest one simple
thing:
  the observed $f_0(600)$ state is the $\sigma$ meson
being responsible for the spontaneous breaking of chiral symmetry.
In 1+1 dimensions, using an exact solution for the large-N limit,
Ref.~\cite{shrock} showed dynamical generation of the sigma meson
starting with the nonlinear sigma model. In 4 dimensions the
situation remains to be  understood. In any case in the linear
$\sigma$ model the $\sigma$ meson is the chiral partner of the
pseudo-goldstone bosons and the two mix each other under chiral
rotations. It suggests that the $\sigma$ meson shares one dynamical
property of the pion, i.e., being a collective mode. Hence the
appropriate way to describe it is not to consider it as dynamically
generated from pion fields, rather it should be discussed at the
equal level as pions in the effective lagrangian approach. Also the
$\sigma$ meson should not be properly described by any simple quark
models since a pion can not be.

 This
work is supported in part by National Natural Science Foundation of
China under contract number
 10575002 and 
 10421503.


\newpage
\begin{figure}[h]
\begin{center}
\mbox{\epsfxsize=150mm\epsffile{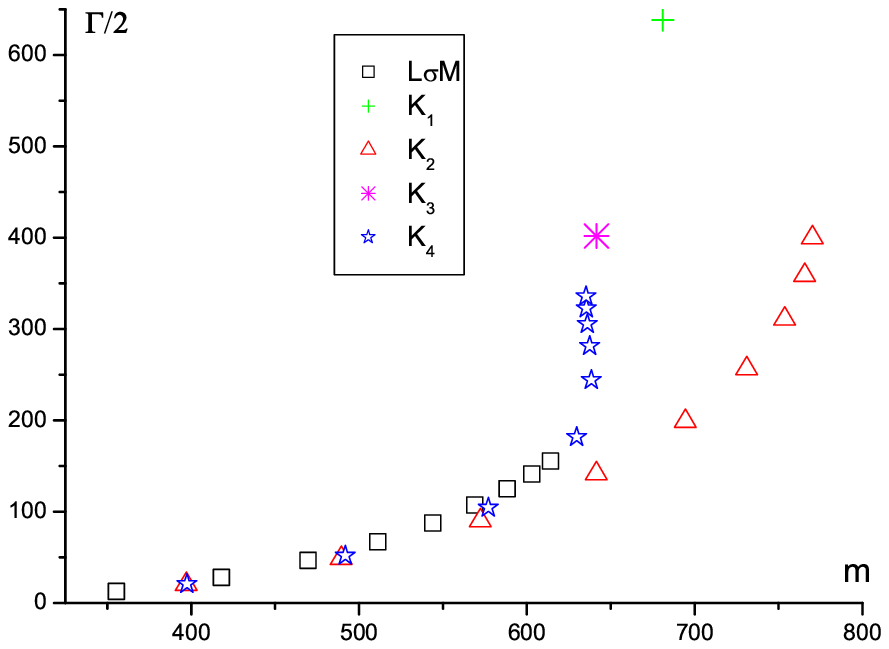}}%
\caption{\label{fig1}The $\sigma$ pole position from different $K$
matrix unitarizations comparing with the rigorous solution. The open
square denotes the exact position of the $O(N)$ model, others are
explained in the text. }
\end{center}
\end{figure}

\begin{figure}[h]
\begin{center}
\mbox{\epsfxsize=150mm\epsffile{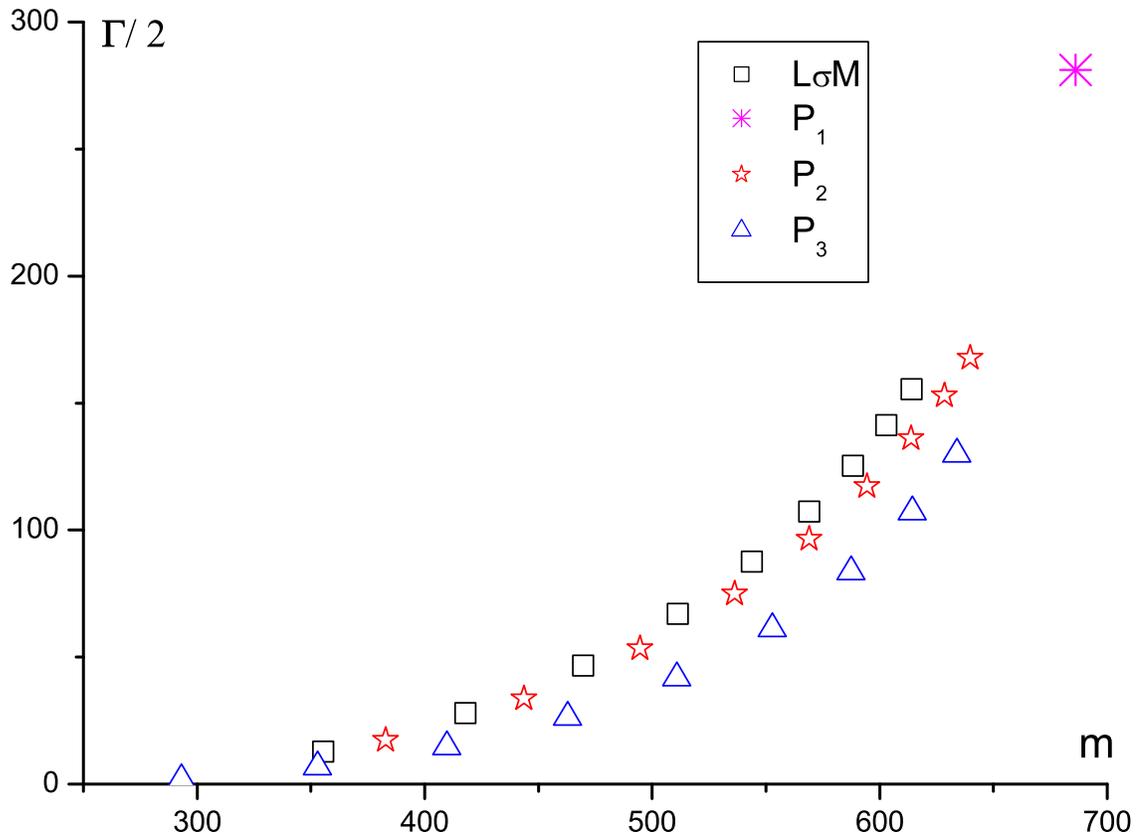}}%
\caption{\label{fig2} The $\sigma$ pole position from different
Pad\'e approximations comparing with the rigorous solution. The open
square denotes the exact position of the $O(N)$ model, others are
explained in the text.  }
\end{center}
\end{figure}
\begin{figure}[h]
\begin{center}
\mbox{\epsfxsize=150mm\epsffile{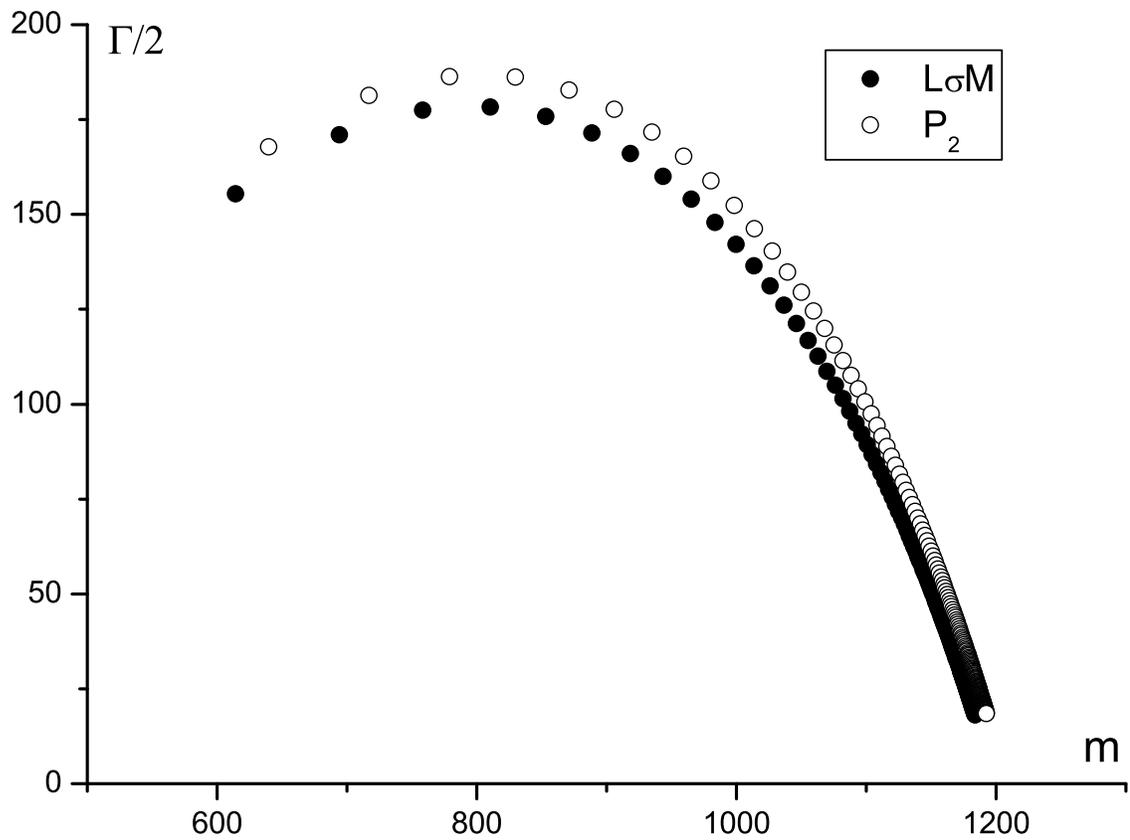}}%
\caption{\label{fig3} $N_c$ dependence of the $\sigma$ pole
position. Full circle: linear $O(N)$ sigma model; open circle: $P_2$
approximation. In drawing the plot $N_c$ starts from 3 and each pole
is differed by increasing $N_c$ by 1. The bare mass $m_\sigma$ is
taken as 1.2GeV. }
\end{center}
\end{figure}
\end{document}